\documentclass[conference,a4paper]{IEEEtran}
\usepackage[T1]{fontenc}
\usepackage[utf8]{inputenc}
\usepackage{mathptmx}
\usepackage[scaled=0.90]{helvet}
\usepackage{courier}
\usepackage{microtype}
\usepackage{graphicx}
\usepackage{booktabs}
\usepackage{tabularx}
\usepackage{array}
\usepackage{xcolor}
\usepackage{listings}
\usepackage{algorithm}
\usepackage{enumitem}\usepackage{float}
\usepackage[hidelinks]{hyperref}
\usepackage{xurl}
\usepackage[numbers,sort]{natbib}

\setlist{itemsep=2pt,topsep=4pt}
\floatname{algorithm}{Algorithm}
\let\oldtexttt\texttt\renewcommand{\texttt}[1]{\oldtexttt{\frenchspacing #1}}
\usepackage{dblfloatfix}
\usepackage{flushend}

\usepackage{etoolbox}
\AtBeginEnvironment{table*}{}
\renewcommand{\paragraph}[1]{\par\smallskip\noindent\textbf{#1}}
\title{Adversary Brains: Hypothesis-Based Logical Signatures for Threat Actor Detection with LLMs}
\author{\IEEEauthorblockN{Gadi Evron}\IEEEauthorblockA{\href{mailto:gadi@knostic.ai}{\texttt{gadi@knostic.ai}}}}

\begin{document}
\maketitle
\suppressfloats[t]

\section*{Abstract}
An adversary brain is an LLM-based model of a threat actor’s behavior, held as an evolving logical signature: a hypothesis with enough structure that data can test it continuously, without a person re-interpreting it each time. Rather than any single observation, such as an IoC (Indicator of Compromise), it is the arrangement that becomes a detection, or intelligence, mechanism. Where YARA matches a file and Sigma matches a log line, an adversary brain matches the activity pattern of a campaign or threat actor. This is based on an intelligence analysis technique called Analysis of Competing Hypotheses (ACH), which the design lifts from the limitation of a few hypotheses at a time, to thousands, and then also operationalizes it for continuous detection purposes. Language models make the concept, and scale, practical: they turn prose reports into brains, judge matches, and drive the loop that keeps brains current. A partial pattern in a brain can also match a new value, such as a fresh C2 (command and control) domain name, assisting with the discovery of new indicators, and turning unknown unknowns into known unknowns, and potentially into knowns.

\medskip
\noindent The author would like to thank Ryan Moon for the collaboration, and Sounil Yu, Chris Inglis, Thomas Roccia, Marius Benthin, Israel Grimberg, Michal Kamensky, Justin Borland, Brandon Levene, John Hultquist, and John Bambenek for their feedback and assistance.

\section{Introduction}

The core idea is that a threat actor’s activity becomes detectable through how its behavior aligns against an \textbf{adversary brain} (a “brain” for short), a structure that represents a logical signature of their operation.

As data comes in, it is matched against every brain at once: it sticks to the brains it fits and falls through as noise where it fits none.

\begin{figure}[!b]
\centering
\includegraphics[width=\columnwidth]{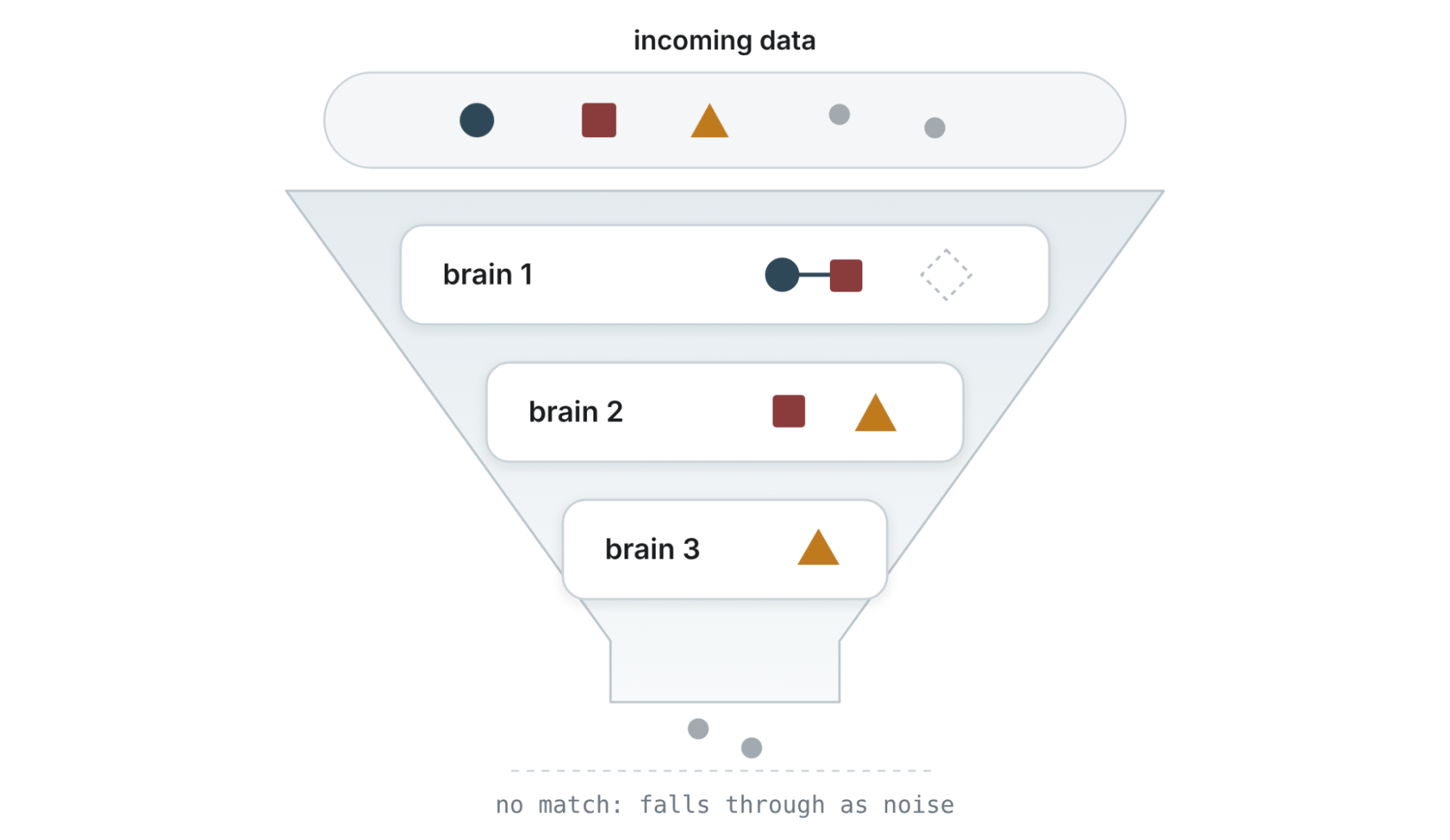}
\caption{Each brain catches the observations that fit its slots, single shapes are entities and a joined pair is a link, open slots wait for the rest, and whatever fits no brain falls through as noise.}
\label{fig:1}
\end{figure}

Rather than any single observation, such as an indicator of compromise (IoC), it is the arrangement that becomes a detection, or an intelligence, mechanism. Where a YARA rule matches a file and a Sigma rule matches a log line, an adversary brain matches the campaign and the threat actor themselves, which live across time, systems, and kinds of data, potentially across multiple artifacts.

\begin{figure}[!b]
\centering
\includegraphics[width=\columnwidth]{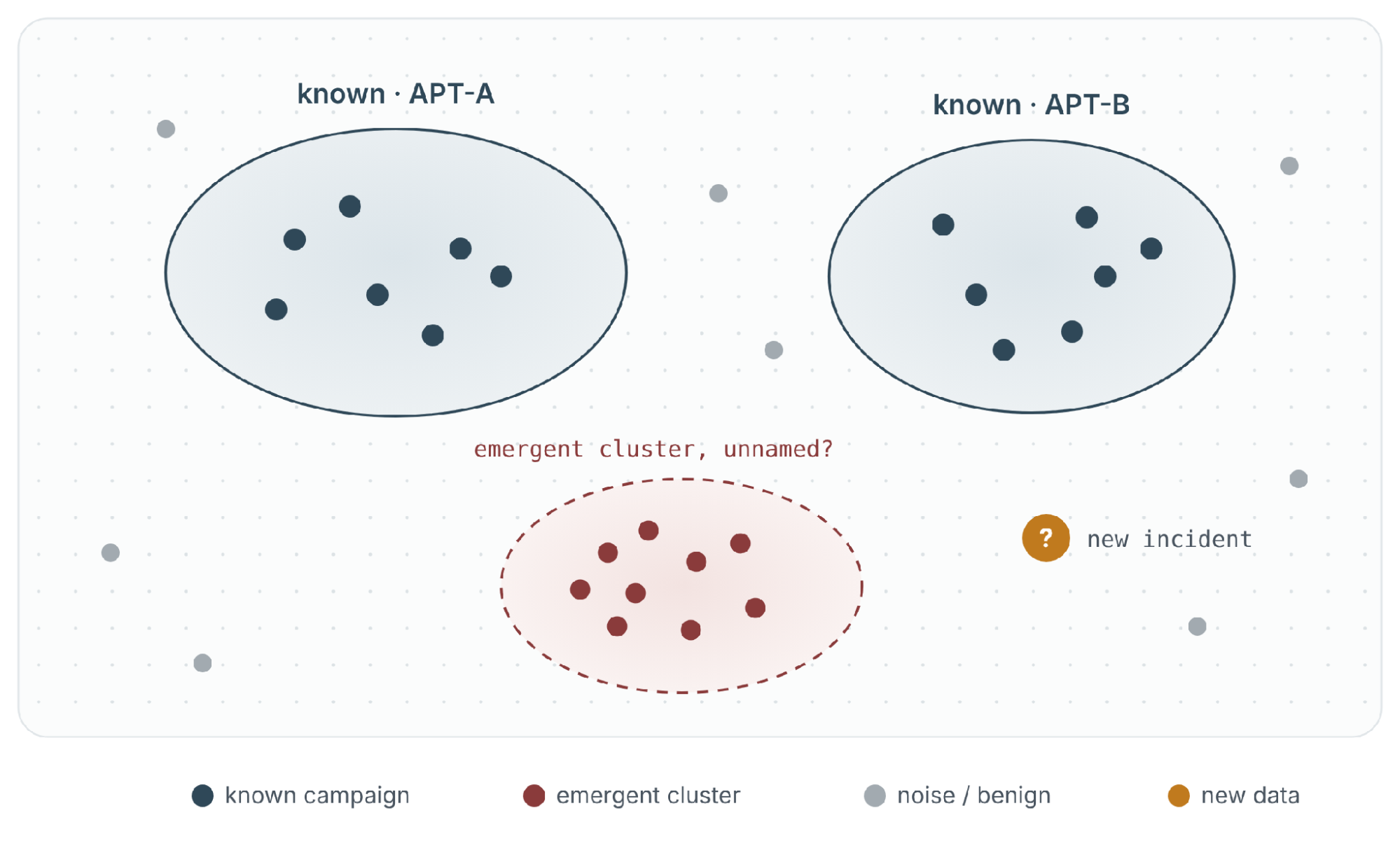}
\caption{A behavioral clustering, where known campaigns and threat actors form labeled regions (in our case, brains), a dense unlabeled group is a candidate new brain, scattered points are noise, and new data, or incidents, are tested against all of them.}
\label{fig:2}
\end{figure}

A brain is the geometric structure over the state and activity signals, held primarily as a knowledge graph that a campaign can be matched against. State and activity are worth pinning down first:

\begin{itemize}

\item \textbf{State}: what is. A standing condition, either one entity, like a hash or an IP, or a relation between entities, like two domains sharing a registrant.

\item \textbf{Activity}: what happens. An event, like a process spawning a child that executes a command.

\end{itemize}

Detection fires on activity only. By analogy, three orbs sitting in space won’t trigger an alert unless a collision risk is detected. An IoC is state, which provides the most context when read with the activity related to it.

State and activity are frames we place on observations so they can be stored and matched. Some inputs can be established as a relation when they arrive. A shared registrant can be read as state (two domains that are related), or deduced as activity (the registration that linked them). Each observation is committed to a frame.

\begin{figure}[!tb]
\centering
\includegraphics[width=\columnwidth]{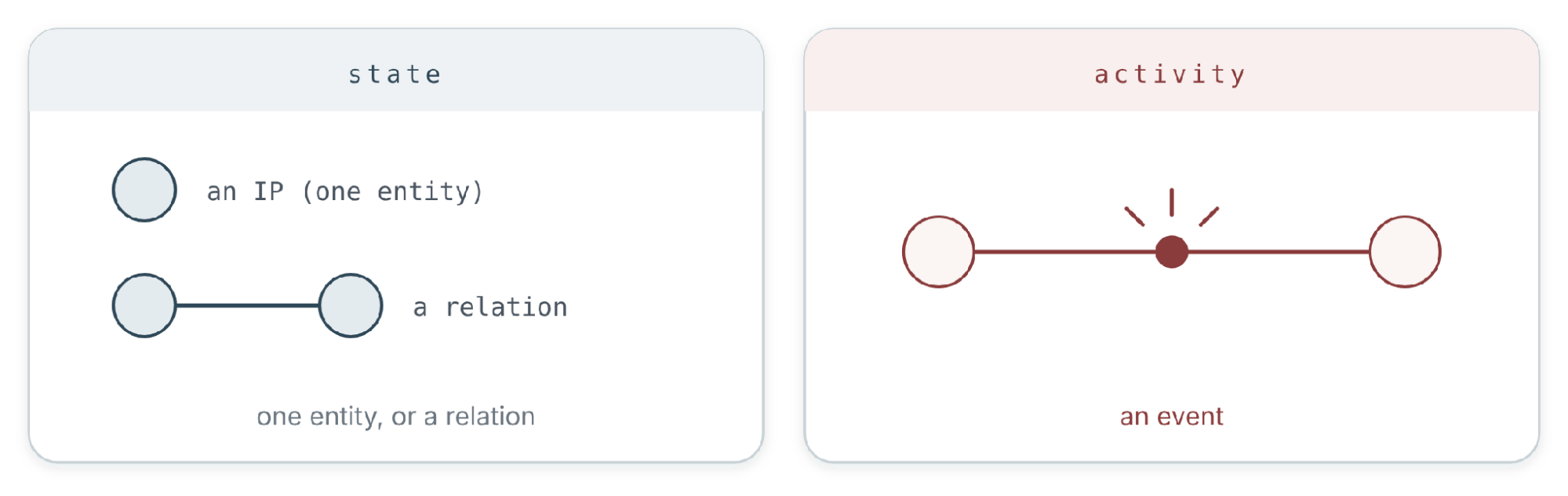}
\caption{State is what is: one entity, or a relation between entities. Activity is what happens: an event. Depending on whether it is provided data, or a deduction, a shared registrant can be read either way, as state or as the registration that produced it.}
\label{fig:3}
\end{figure}

Today knowledge of how an actor operates lives in prose such as vendor reports and hand-built detection playbooks, or even as static text a person re-reads and re-applies every time. A brain is the same knowledge made into a logical signature. Data tests it directly, and it updates itself as new observations and new hypotheses arrive. (IoCs are also cheap for a threat actor to swap while these patterns could take effort to change.)

\paragraph{Goal.} This text proposes a more AI-native way to do threat intelligence: running its claims as live hypotheses, with language models to make that practical at scale. Analysts already turn some claims about how an actor operates into detection rules. The rule may cite the claim, and its fires do not flow back to revise it. Here the claim itself runs against the defender’s own telemetry as data arrives, and its confirmed and dismissed fires raise or lower its standing, and can retire it.

The design has three aims. The first is to hold thousands of such hypotheses at once. The second is to make TTPs (tactics, techniques, and procedures) detectable as specific arrangements of observations, mostly at the level of procedures. The third is to turn unknown activity into known unknowns, and potentially, some of those into knowns~\cite{r1}\cite{r2}.

\paragraph{Why build brains.} The first four uses below help run a security program. The last two work inside detection.

\begin{itemize}

\item \textbf{Where to invest.} To start, a cheap test on past data suggests which actors are worth a brain.

\item \textbf{Lasting effort.} A brain marks which of its links are expected to last, so effort can go to those.

\item \textbf{Named gaps.} A brain that matches only in part names the missing evidence. A step the sensors cannot see points to telemetry to add.

\item \textbf{Open assessments.} A brain stays a hypothesis that data and analyst verdicts can confirm or retire.

\item \textbf{Arrangements.} Where every step by itself is common, steps taken in order on the same account or host can still be the signature.

\item \textbf{Graded use.} By design, a brain alerts, ranks a queue, or seeds a hunt, depending on how often its matches are confirmed.

\end{itemize}

\paragraph{From intelligence analysis to detection.} Analysis of Competing Hypotheses (ACH) is a technique developed at the CIA for intelligence and national-security analysis~\cite{r3}. The design takes it to thousands of hypotheses, and then from an analyst’s assessment to operational detection.

In ACH, an analyst scores several explanations against the same evidence and gives most weight to the evidence that rules explanations out. It is done by hand in a matrix. Heuer notes that “more than seven hypotheses may be unmanageable”~\cite{r3}.

Each hypothesis becomes a brain, a machine-checkable object. A system can then hold thousands of them and score every one against the same incoming data, continuously.

That scoring runs inside detection, next to signatures such as YARA and Sigma. An IoC becomes one link in a hypothesis, and other IoCs can fill the same place. The analyst moves from filling in the matrix to judging its output: confirming or dismissing fires, and deciding drift and attribution.

All of this rests on one bet: that a threat actor’s activity structures remain somewhat stable, and rare enough to be worth representing as objects, instead of detecting specific descriptors of behaviors one at a time. If it holds, brains are worth their complexity.

Operationally, such structures sometimes don’t remain stable for long. Many actors shift their m.o., with a few durable pieces staying stable in the pattern. Depending on the delta, in such cases, a simpler detection and enrichment stack achieves the same or better results. The test in “Whether to build, and how” is how you reach that verdict cheaply, before over-building.

We built a reference implementation of the schema in appendix 1 to test the schema’s rules (appendix 3).

\paragraph{Threat model.} The adversary is the actor being detected, which rotates cheap indicators freely. When extracting brain data from threat reports, these can also potentially be seeded with misleading structures. A brain is therefore born unconfirmed and must survive testing on actual telemetry before it is trusted. Then, as the defender’s telemetry also reflects the adversary’s own actions, a model judging a match can be potentially fed activity constructed to mislead it. The deterministic screen and the rule that model evidence is drawn only from the deterministic match limit these risks.

\paragraph{What this text establishes, and what it does not.} This is a design and a specification, with a reference implementation. It works through how public reports populate into brains and how a cheap test separates actors worth representing from actors that are not.

The text does not measure detection on real telemetry: no detection rate, false-positive rate, or robustness under indicator rotation against a signature baseline. Producing those numbers is out of scope. The figures in this text are schematic. The measurements in appendix 3 come from synthetic data, one emulated case, and later public reports. While the system was run with limited cases in live environments, that data wasn’t taken into consideration for this version of the text. Matching cost at the thousands-of-brains scale the design assumes is not analyzed here.

The repo for the Adversary Brains reference implementation (appendix 3) can be found here: \url{https://github.com/gadievron/adversary-brains}.

\section{What a brain is}

A brain is a \textbf{knowledge structure}. A database records what happened. A brain records what it would \emph{mean} if something happened: “if these observations show up, arranged roughly like this, that is evidence for a trigger, and for the brain’s validity.” The \emph{if \ldots{} then it means \ldots{}} is what makes it a hypothesis rather than a log, and giving it structure is what lets a machine test it.

A brain is built from two kinds of piece: \textbf{entities}, the observables like a hash, an IP, or a process, and \textbf{links}, the typed relations between them (this one process started that one, two domains share a registrant). A single entity can match on its own, by its value or by a pattern. What makes either one strong is rarity. Even at the single indicator level, a rare hash places an actor well. A common IP does not.

Each piece has a \textbf{weight}: a number for how much a match on that piece counts toward the brain. It depends on how often the piece shows up in ordinary traffic. A file hash that never appears in your normal traffic gets a high weight, but a cloud provider’s IP address that thousands of your hosts contact every day gets a low one. Each defender computes weights from their own traffic (formally, the negative log of the piece’s background rate). Because weight comes only from background traffic, it stays the same when brains are added, split, or merged.

Not every link must hit. A brain marks some links required and others supporting. It can fire once its required links are met and enough of its supporting links match.

Beyond weight and type, a link carries a role (required or supporting), whether our sensors can see it (its observability), and a volatility, meaning how fast it is expected to change. A code-reuse signature is stable and counts for longer, whereas a URL is volatile, expected to rotate, and fades on its own.

Volatility tends to track the Pyramid of Pain~\cite{r4}: atomic pieces at the bottom rotate fastest and fade soonest, while behavioral links higher up change slowly, so a brain persists on its durable upper links while its lower ones churn. Volatility is the author’s forecast and can be wrong: a link tied to one platform looks stable until the actor moves to another. Together, observability and volatility mark which links may be invisible here and which are expected to drift.

A brain can rest on one distinctive piece more than on many indicators. When Mandiant exposed APT1 in 2013 and published over 3,000 indicators, the group stopped using most of that infrastructure within about three months~\cite{r5}. The habit behind its “Comment Crew” nickname, hiding commands in web-page comments~\cite{r6}, is a behavior, which costs an actor more to change~\cite{r4}. Appendix 2 builds this brain in full.

Each brain has two settings, chosen independently. Its representation is how its content is stored: a list, a chain, a centroid, an embedding, or a graph. Its match method is how new data is scored against it: exact match, distance, similarity, sequence, or likelihood (how probable the observations are if the brain’s pattern is real).

Both are set in the brain’s definition, so the engine does not force one geometry on every brain. For each brain it checks whether the data fits, and how well.

The default is a knowledge graph, since most brain content is relational and state-heavy, which a graph represents best. The other forms fit better for different types of content.

A chain fits steps whose order is the signal, such as SUNBURST’s signed Orion update, then a dormant period of up to two weeks, then DNS lookups to its command server~\cite{r7}. A list fits a set that matters as a whole, such as the file types Akira’s encryptor targets: databases, virtual memory, and disk images, while it leaves PDFs and Office files alone~\cite{r8}. A centroid or an embedding fits a brain learned from the data, with its confirmed matches, with no written rules.

\begin{figure}[!tb]
\centering
\includegraphics[width=\columnwidth]{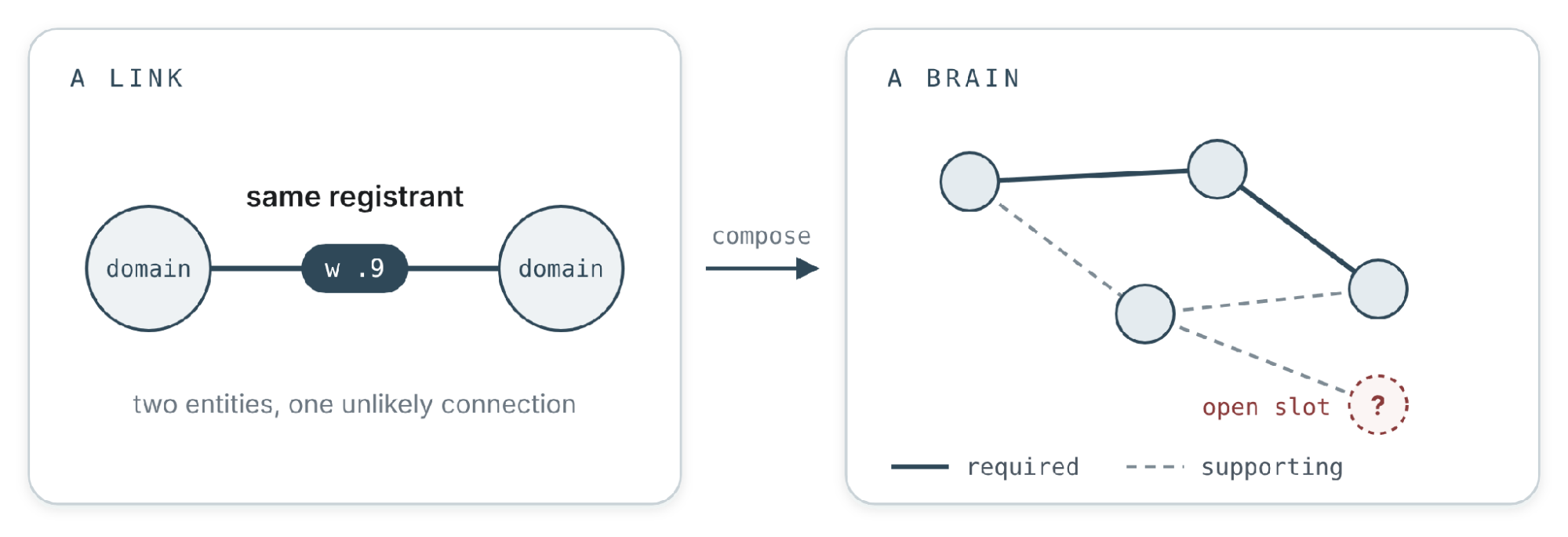}
\caption{Links compose into a brain. Solid links are the required core it cannot fire without. Dashed links are supporting. Open slots wait for what completes them.}
\label{fig:4}
\end{figure}

A bundle of these entities and links is a brain, and it works at any size. It is the same object whether it has two links or two hundred. It is also \textbf{disposable}: never simply true or false, only alive or dead. Brains can be retired, although in theory we can run thousands and lose nothing if they never trigger.

\paragraph{This is how a brain differs from a TTP.} A technique such as “credential dumping” is a \emph{label} for a kind of behavior. On its own it fits many unrelated actors and asserts nothing. A brain is a \emph{claim}: a specific weighted arrangement of observations, sometimes with techniques among its pieces, that data can build up or tear down.

In MITRE ATT\&CK’s terms, tactics are the why and techniques are the how. Procedures are “the specific implementation the adversary uses” for a technique~\cite{r9}. Most entities in the reference implementation’s brains are concrete observables, such as a process name or a command line. So while not limited to it, what a brain holds sits mostly at the level of procedures, whatever its scale. With that said, some of the study brains in appendix 3 hold only state, such as a shared certificate or a build path, and so cannot fire under the rule in appendix 1.

The schema has no dedicated field for tactics. A brain can still carry them, as notes on its entities or as a custom entity type, and an attack chain orders its steps much as tactics do. A tactic is a goal, so it can match only where telemetry already labels events with one, such as alerts tagged with ATT\&CK tactics.

\paragraph{The same object, nested.} Each level is built from the ones beneath it, which is what \emph{scale-invariant} means in practice.

\begin{enumerate}

\item \textbf{A link.} Two domains sharing a registrant (state), or a login then a privilege escalation (activity). The smallest unit, an entity or a relation between two.

\item \textbf{A technique set.} A few weighted behaviors. MITRE ATT\&CK IDs are one source~\cite{r9}, though not the only one. On its own it discriminates poorly, since common techniques are shared by everyone. A set like this is worth keeping only when the arrangement is rare while the pieces are common, the way one sequence of an MFA reset, then a new identity provider, then a loosened access policy is distinctive, though each step alone is routine.

\item \textbf{An attack chain.} The above techniques and links in a specific arrangement across one intrusion, the idea behind the intrusion kill chain~\cite{r10}. For example: lure $\rightarrow$ macro $\rightarrow$ downloaded second stage $\rightarrow$ persistence $\rightarrow$ credential theft $\rightarrow$ lateral movement.

\item \textbf{A campaign pattern.} The shape shared by several attack chains that work together.

\item \textbf{A threat actor.} Many campaign patterns under one operator.

\end{enumerate}

Move up the ladder and each brain contains all or part of the ones below it, so an actor-scale brain is composed of many links, with mixed durability, firing on whatever subset is present. Where an actor’s campaigns share no common core, it may not reduce to one brain at all (appendix 1).

(A brain need not enumerate links at all. It can be a region in a behavioral “space” where fit just means landing nearby. Such a brain is a region learned from its confirmed matches, a centroid or a boundary, and fit is distance to that region. The space and the distance can be a learned vector embedding or a simpler measure of shared features. The brain’s representation and match method say which. The ladder is illustrative.)

\section{How it works}

The system runs ACH, introduced above, by machine. The moving parts are few: the brains, a pass that scores incoming data against every brain, and a loop that reruns the pass continuously and turns confirmed and dismissed matches into updates.

Observations enter through ingestion and normalization, which appendix 1 details. The pass and the operations of a later section are one machinery seen twice: Match is the pass run forward, Discover carries its new-brain outcome forward, Extract enters from a source rather than from the stream, and Coverage-map, Compile, and Share connect a brain outward to your stack and peers.

The system runs autonomously on the brains it is fed, a human optional throughout. It does not, however, make these two judgments on its own: attribution, and confirming that a brain has drifted.

Clustering and rarity weighting are long established in threat intelligence~\cite{r11}\cite{r12}\cite{r13}, and multi-event rules such as Sigma correlations, EQL sequences, and risk-based alerting already join events in detection~\cite{r14}\cite{r15}\cite{r16}.

This is similar to how a Sigma correlation can require several rules to match within a timespan, grouped by a shared field such as one account~\cite{r14}. Its temporal\_ordered form also requires order. Few Sigma backends implement that form, and EQL sequences do it natively~\cite{r15}. Risk-based alerting sums the risk scores that rules assign to each entity over a period~\cite{r16}.

Attack Flow describes how an actor’s actions connect, as an extension to STIX (Structured Threat Information Expression) 2.1, with AND and OR operators~\cite{r17}. Its measure of belief is the optional STIX confidence, a 0 to 100 score its author can set on each object. A brain adds weights computed from each defender’s own background rates, suspension when a required link cannot be seen, and partial matches that name the links still missing. The weights and suspension apply only to the unordered fire rule. That rule has no OR over required links, which Attack Flow’s OR operator allows (appendix 1).

\paragraph{The innovation.} One object is authored as intelligence, compiled to a running detector, and is revised by that detector’s own confirmed and dismissed fires, with attribution held out by construction.

Building attack graphs from host telemetry is proven ground. POIROT aligns graphs built from threat intelligence against kernel audit records~\cite{r18}. HOLMES correlates suspicious information flows into a high-level graph of the attack~\cite{r19}. UNICORN and Kairos flag departures from learned system behavior without prior attack knowledge~\cite{r20}\cite{r21}. What a brain adds is the lifecycle: the graph is held as a competing, revisable hypothesis rather than a fixed signature to search for.

\paragraph{Many hypotheses, held live at once.} Thousands of competing brains run side by side, and each incoming observation is scored against all of them in one cheap pass. That pass does four things at once. It \textbf{corroborates} (an observation strengthens a brain it fits), \textbf{detects} (enough pieces line up that a brain completes, and we alert), \textbf{screens} (an observation that \emph{contradicts} a brain weakens or rules it out, and only a contradiction counts, never a mere absence, since not seeing something may only mean we are blind to it), and \textbf{generates} (an observation matching nothing existing starts a new brain to watch).

\begin{figure}[!tb]
\centering
\includegraphics[width=\columnwidth]{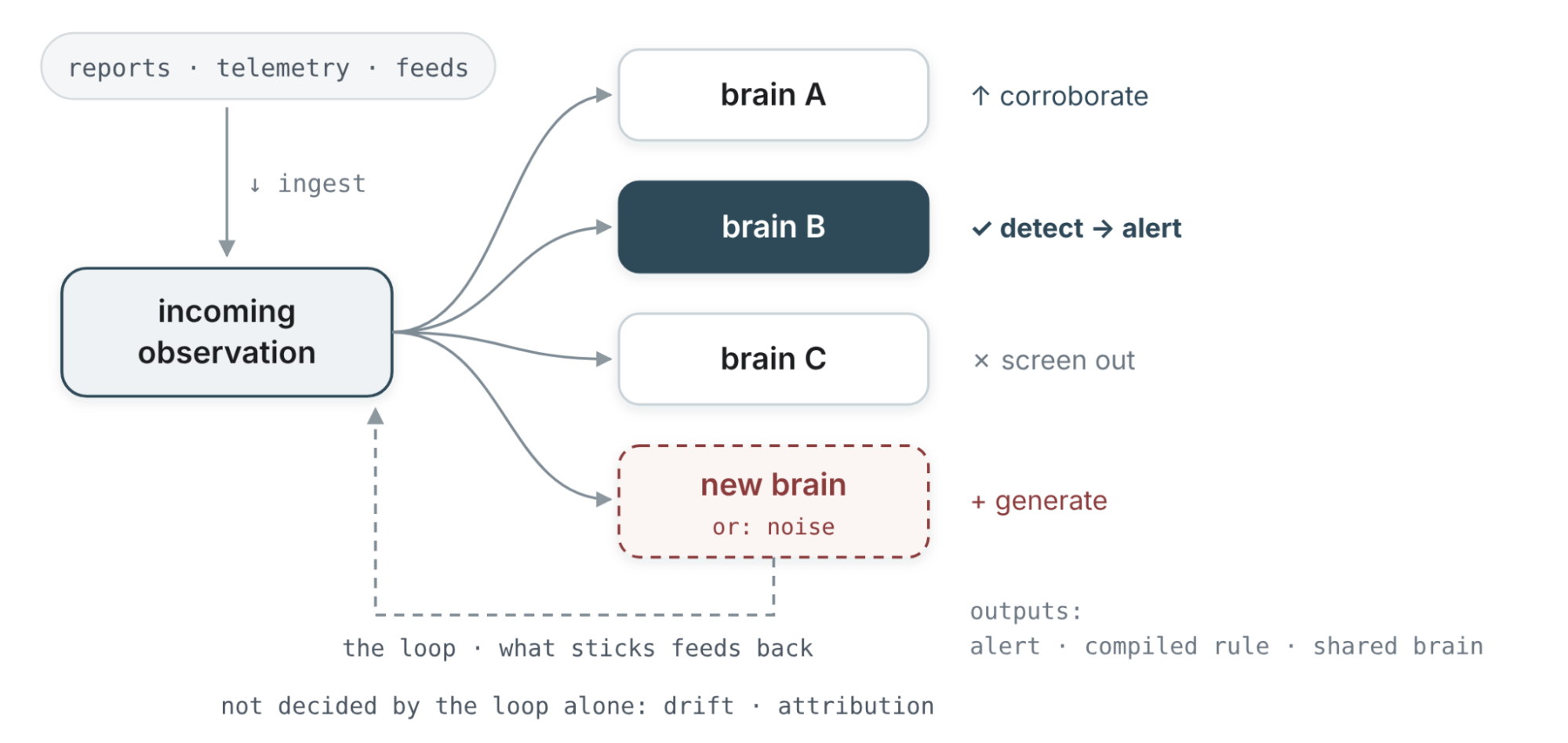}
\caption{One pass, many live hypotheses: the same scoring both matches known brains and seeds unknown ones, and leaves the rest as noise. The result feeds back into the loop. Ingestion feeds the pass. Alerts, compiled rules, and shared brains leave it, and the drift and attribution judgments are not made by the loop alone.}
\label{fig:5}
\end{figure}

Grouping observations that fit no brain into a new one is clustering. Declining to force such an observation into the nearest existing brain is open-set recognition~\cite{r22}. A brain born this way from data is an emergent cluster. Mandiant’s uncategorized (UNC) groups~\cite{r11} and the Diamond Model’s activity groups~\cite{r12} form clusters the same way, for questions such as who is behind an intrusion. Here the cluster is used for detection.

\paragraph{Every scale at once.} Because a brain is scale-invariant, one pass scores an observation against brains of every size in the same sweep: a technique-level brain, a campaign-level brain, an actor-level brain. It matches wherever it fits, which may be several scales or none. A scale with no matching brain is passed over without counting as a failure, and may be where a new brain later forms.

\begin{figure}[!tb]
\centering
\includegraphics[width=\columnwidth]{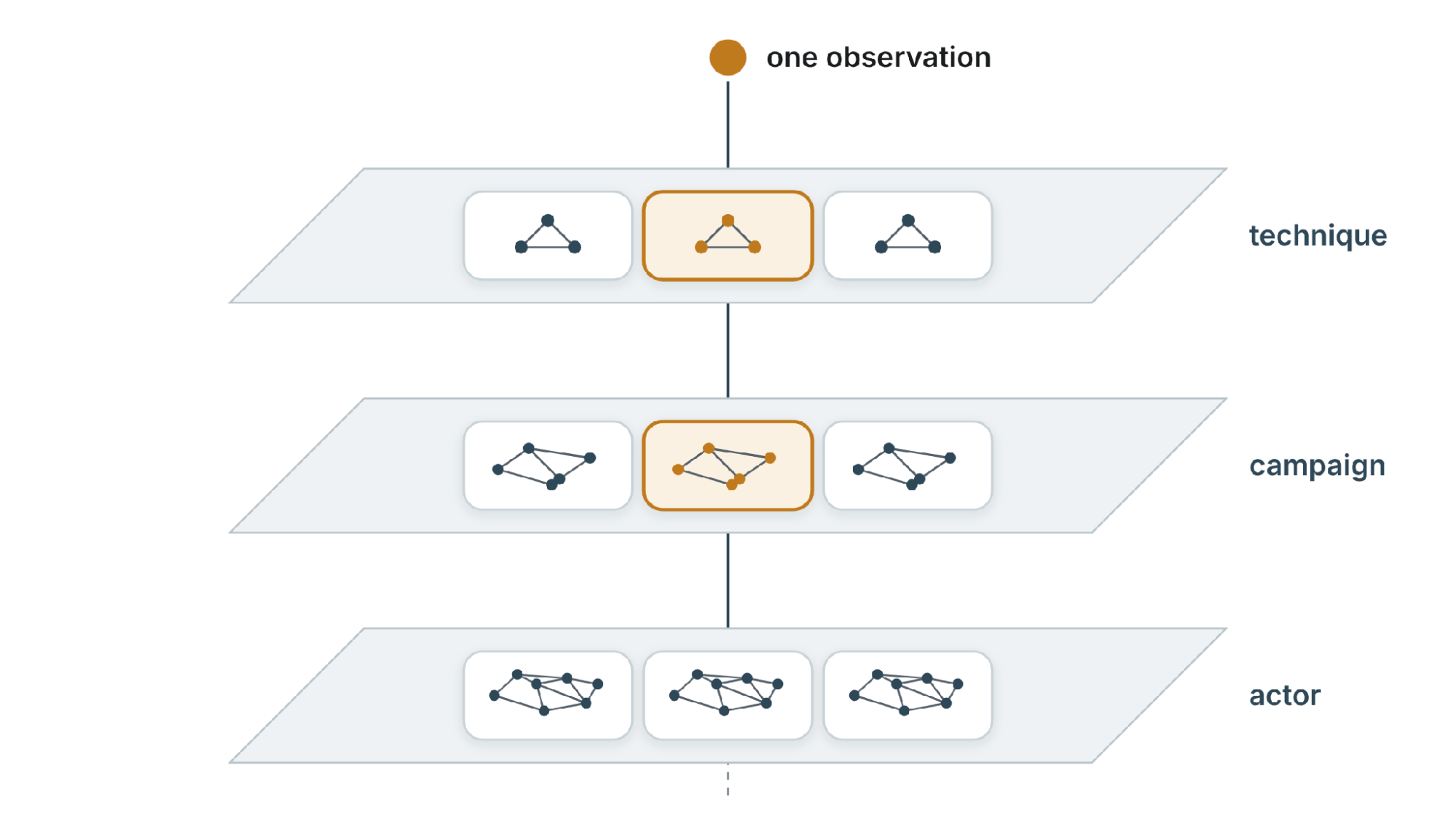}
\caption{One observation, scored against brains at every scale at once. Here it matches a technique brain and a campaign brain, and passes through the actor scale, where nothing fits yet.}
\label{fig:6}
\end{figure}

\paragraph{The questions.} Underneath, a brain can be “asked” four questions, broad to narrow.

\begin{enumerate}

\item \textbf{What sticks?} Of the whole stream, which observations attach to this brain at all.

\item \textbf{Does this one observation fit, and how well?} The single-point score the cheap pass produces.

\item \textbf{Do these observations together fit: all of the brain, most, or a part?} Real intrusions could arrive as partial fits that accumulate.

\item \textbf{What would complete you?} Run backwards: a half-finished brain names the evidence worth searching for, making it an actionable hypothesis, from a filter into a collection task, the collection-and-analysis turn of the intelligence cycle run by the loop with no analyst tasking it~\cite{r23}. The missing evidence becomes a known unknown. It can become a known when collection finds it or rules it out.

\end{enumerate}

\paragraph{The loop.} This runs continuously: each pass scores incoming data against the brains and updates each brain’s standing, a scalar prior raised by corroborating evidence and lowered by contradicting evidence, with the size of the move set by the evidence’s weight.

An absence does not move it, since it may only mean we lack data. On that standing the loop splits a brain that starts catching two patterns, merges two that turn out to be one, fades one that stops matching, and retires one whose matches are never confirmed.

It works to disprove its leading brains before agreeing with them, since a loop that only seeks confirmation reinforces its own “beliefs.” For example, should it trust rare links more than common ones, as discussed earlier, and under what circumstances?

Confirming drift is a judgment the loop does not make on its own. A brain’s matches can move away from what it used to catch for three reasons: the actor changed, a new pattern split off, or two actors were fused together. Only the first is concept drift~\cite{r24}.

\paragraph{An agentic loop.} This is designed as an agentic loop in the term’s current sense: a model-driven loop that directs its own process and chooses its next action~\cite{r25}. This one also runs unattended and holds persistent objects. Designing such a loop, rather than prompting the agent, is what practitioners now call loop engineering~\cite{r26}.

The loop sits between two familiar loops and differs from each. Unlike a detection loop, whose rules stay fixed while analysts tune them, it rewrites its own objects, splitting, merging, and retiring brains.

Unlike the build and research loops that rewrite their working state each pass, Karpathy’s autoresearch and Huntley’s ralph among them~\cite{r27}\cite{r28}, its outputs never re-enter as evidence: those loops accumulate what each pass writes, with a test or a metric as the verifier. Here the verifier is telemetry, the objects are competing hypotheses, and a hypothesis fed by its own conclusions is barred by the schema, the self-reinforcement risk evolving-memory systems must govern~\cite{r29}.

Two practical notes. First, the data a brain is matched against arrives in different shapes: relational content sits in the graph natively, while event sequences (network logs, for instance) are time-series and fit a graph poorly, so matching a temporal pattern reaches into time-series data. A brain spanning both is matched piece by piece and rejoined.

Second, comparing scores across brains only matters when they share one ranked list. The shared quantity is a p-value: the probability of a match at least this strong under a null model in which the brain’s pieces appear independently at their measured background rates. Cross-brain comparability of these p-values, across brains using different match methods, is a limitation not treated by this text.

Where the system uses language models, it constrains them. Language models turn prose reports into brains and judge whether events match. Embeddings give brains that need no written rules. Extraction is bounded by validation against canonical sources, and every model verdict must cite evidence from the deterministic match block, so a hallucinated technique id is dropped and ungrounded evidence is discarded (appendix 3). One thing this does not fix: a model’s confidence score is not calibrated. We use it only to rank.

Three parts of the system also cover for each other. Discover proposes brains from data, the competing-hypotheses scoring tries to disprove them, and the loop drives the collection that settles which survive.

One note on the name. In agent systems, a brain is the reasoning core that runs a loop. Here a brain is the object that gets matched against, a model of the threat actor’s behavior. It describes what they do, and says nothing about how they think. The agency in this system sits in the loop around the brains, never in a brain itself.

\paragraph{What a brain is good for.} Appendix 2 applies the test from “Whether to build, and how”. As a negative example, it finds no actor-scale brain worth building for the ransomware floor. For Scattered Spider’s phone-call step, it points to better help-desk verification.

Volatility marks which links are expected to last. Later reports bore this out in part, where all three rotating links they covered being changed. Two Scattered Spider links authored as stable, both tied to Azure, were replaced by VMware techniques. That forecast missed (appendix 3).

Matching can turn unknown activity into known unknowns. An entity can be a pattern with no fixed value, so a value never seen before can fill it. In a synthetic test in appendix 3, a behavioral brain kept firing on all 99 runs with rotated indicators.

When the rest of an arrangement holds and one step stops matching, the step may have changed, been missed, or been skipped, each pointing to a different check. Activity that fits no brain stays unknown, unless it recurs often enough for Discover to propose a brain for it. These proposals would likely include benign patterns that still need pruning.

Volt Typhoon leaves few distinctive artifacts, and two of its native actions fit a small brain. Scattered Spider’s identity steps are the case where the arrangement is the signature (appendix 2). Appendix 3 tests ordered steps by one user on synthetic data.

Within a detection stack, a brain carries a fidelity, its confirmed-match rate, and that score sets its use. A specific, high-fidelity brain can alert on its own. A commodity or low-fidelity brain does not alert alone. It raises the confidence of brains it fires alongside, ranks a triage queue, or seeds a hunt. Below the alert line it adds ranking signal without raising alerts, and a larger brain fires once where its pieces would each fire. A generic brain that corroborates nothing and is never confirmed loses fidelity until the loop retires it.

\paragraph{What “logical signature” means.} A logical signature describes an arrangement: the pieces and how they must relate, with specific values optional. A brain keeps what makes a signature useful: it is declarative, portable, and compiles to something a stack runs cheaply. It differs in staying revisable, and the loop reads decay in its matches as drift.

By design, a multi-link brain degrades rather than collapses when one link is evaded, though we have tested this only on synthetic data and one emulated case (appendix 3), and have not measured how much resistance it gives. A brain built on a single link is as brittle as a signature. Corroborating links and a running loop are what make it different. Michal Kamensky’s vulnerability variant hunter works the same way~\cite{r30}: a logical vulnerability reduced to its underlying flow and why it can go wrong, with names and literals stripped, holds across languages where a hand-written signature matches only the shape it was given.

\section{What you do with a brain, and what it resolves}

A brain comes into being in one of two ways, and the two cross-pollinate over time. Building a brain bottom-up from data that fits no known patterns is referred to as Discover. Building one top-down, lifted from a report, taken from curated intel, or composed by an analyst from scratch, is Extract. The top-down method provides weighted priors for a detection-ready brain, whereas the bottom-up approach provides observations that confirm or kill the hypothesis they maintain.

The other four operations that use a brain: Match runs it against incoming data, Coverage-map and Compile connect it to your detection stack, and Share passes it to peers. Each operation below names the problem it resolves. Together they keep an assessment from hardening into a fact nobody re-examines: a brain stays a revisable hypothesis, checked continuously against your data.

\begin{itemize}

\item \textbf{Discover.} Unknown campaigns stay invisible until someone writes the report. Discover grows a brain bottom-up from an observation that fits no prior hypothesis pattern, or potentially ties two or more existing brains together.

\item \textbf{Extract.} Threat actor knowledge lives in heads and prose, and every report is re-interpreted by hand. Extract builds a brain top-down from a source: a report, curated intel, or an analyst composing it by hand, holding the knowledge in a form people and machines both write to. The brain is born unconfirmed and testable by your own data.

\item \textbf{Match.} Detection is largely atomic: signatures and rules fire on one artifact or IoC at a time, so an attack spread across many of them is seldom seen whole. Match fans one observation across every brain in one cheap pass, absorbing it into the brains it fits, so the arrangement is what gets detected. Run in reverse to ask what would complete a brain.

\item \textbf{Coverage-map.} Silence is ambiguous: a missing alert can mean the activity was absent, or that your sensors could not see it. Coverage-map projects a brain’s entities and links onto your telemetry and sorts each into covered (a detection exists), visible but uncovered (the data is there, no detection yet), or invisible (no telemetry at all).

\par Observability is whether your sensors can see a link. Coverage is whether a detection exists for it: a Coverage-map result is about your rules, not a value stored on the link. A required link you cannot observe suspends the brain rather than failing it, so a real “no match” means “not there” instead of “not seen.”

\par For some actors the telling behavior sits off your telemetry, a help-desk phone call, a forged token, an encrypted channel, so the brain could suspend because of missing data rather than because the actor is absent.

\item \textbf{Compile.} Matching a full brain is rich but costly. Compile a brain down to whatever your detection stack runs cheaply, a Sigma~\cite{r31} or YARA rule, a SIEM query, a watchlist, a hunt playbook, tuned to fire fast with few false positives. This writes to your stack, so run Coverage-map first.

\item \textbf{Share.} The brain is the shareable object, with its links and structure, verifiable by a peer against their own data. Share those and withhold your attribution: the evidence travels, your conclusion about who is behind it does not have to. Shared intelligence asks the receiver to trust the sender for that conclusion, and conclusions do not transfer between environments.

\end{itemize}

\section{The schema}

The schema is deliberately small: three objects, the entity, the link, and the brain, layered on standards that already exist. It defines no event taxonomy of its own.

Incoming activity is normalized with an open event schema such as OCSF (the Open Cybersecurity Schema Framework)~\cite{r32}, entities are named with STIX vocabulary~\cite{r33}, and the brain schema adds the layer those standards lack: the hypothesis.

Entity patterns are the main place the schema does not reuse a standard. They use a short syntax of its own (appendix 1). The schema bounds what a brain can represent, and so what it can test.

\begin{itemize}

\item \textbf{Link}: two entities joined by one typed relation. It carries the attributes established above (type, role, observability, volatility), an edge weight computed against local base rates, and a provenance: observed in data, derived, or reported.

\item \textbf{Brain}: identity and lifecycle (a stable id and a status), a scale from link up to threat actor, its set of links, its representation and match method (the two settings in “What a brain is”), and lineage: the report it came from or the parents it split or merged from. It also carries a fidelity, its confirmed-match rate, kept locally because it is a record of your own analysts’ verdicts.

\end{itemize}

Entities are the nodes and links are the edges: these two objects are the knowledge graph itself, and a brain’s file is one serialization of that graph, a typed property graph. Node types are drawn from STIX vocabulary, edge types from the link relation set, and edge attributes of weight, role, volatility, and observability. Appendix 1 is its normative definition, and those closed value sets are the minimal ontology.

\paragraph{Enforcement.} Evidence flows only upward: any detection compiled from a brain is tagged with that brain’s id, and ingest drops any observation whose provenance carries such a tag, so a brain’s own compiled detections cannot re-enter as fresh evidence.

This closes the automated feedback path. It does not close the human one, in which an analyst could read a brain’s alert, write it into a report, and that report can later be ingested.

That relay is a residual risk, flagged and reviewed through provenance. The schema enforces a second rule the same way: it has no attribution field, so naming stays decoupled from matching by construction. A brain has no single source to inherit trust from. It gathers links from reports, telemetry, and peers over time. Trust is held per link as provenance and per brain as its fidelity.

\begin{figure}[!tb]
\centering
\includegraphics[width=\columnwidth]{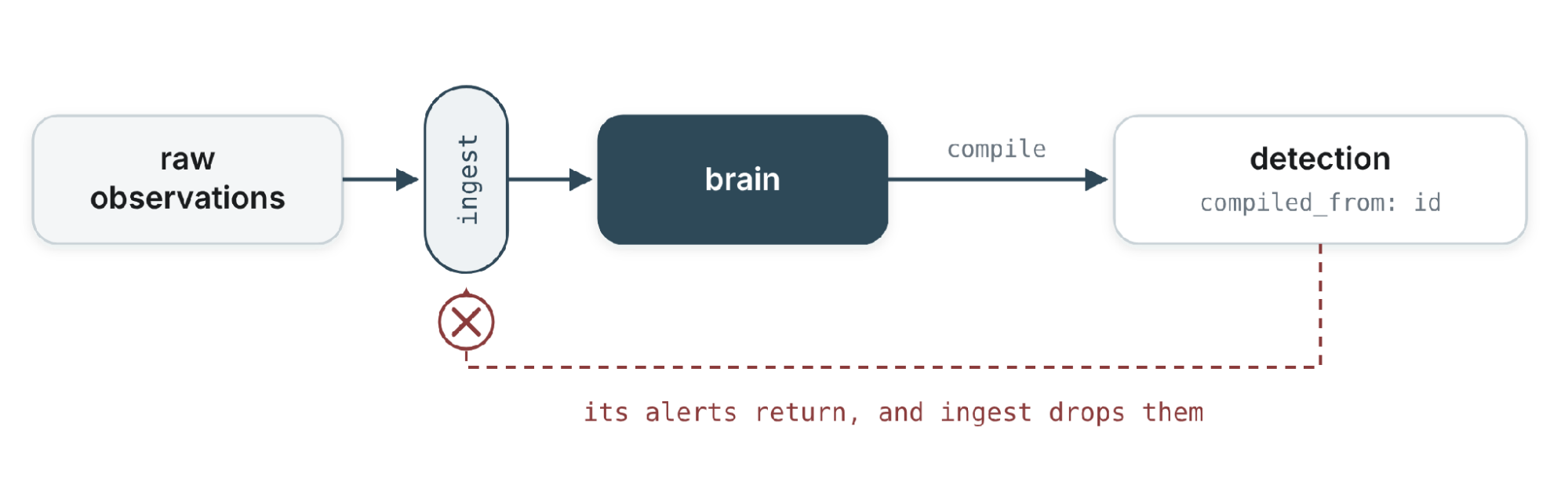}
\caption{Evidence flows only upward: raw observations pass ingest, but an alert from a detection compiled from a brain carries that brain’s id, and ingest drops it. The automated path is closed by construction. The human relay, an alert written into a report and later ingested, remains a flagged residual risk.}
\label{fig:7}
\end{figure}

\paragraph{Sharing, as a repository.} A brain’s definition is a file, and its natural home is a git repository on GitHub, the same shape as the Sigma rule repositories: one brain per file, a stable id, a status, references, and a related field carrying split and merge lineage.

Two rules bound sharing. First, the repository holds definitions, the hypothesis and its links, never live state: weights are computed against local base rates, so they stay home, and each receiver recomputes them against their own background. Second, internal entities are sanitized before publishing, so the links travel without the victim.

The normative specification, version 0.1, is in appendix 1: field names, types, allowed values, sharing rules, file conventions, the repository layout, and a worked example a team can implement from.

\section{Whether to build, and how}

What follows is a staged plan for a minimal first version, testing the idea in your own environment. The first step is deciding whether to build at all.

\paragraph{First, test the wager.} The wager is the bet the whole approach rests on: that this actor left a structure rare, durable, and observable enough to be worth an actor-scale brain over a simpler stack. Before building anything, run a cheap check on data you already have, with no new collection and no attribution. Take a handful of past intrusions you already understand and pull out their characteristic pieces. Then place each one on two axes: how often it \emph{recurs} across related incidents, and how \emph{rare} it is in ordinary background traffic.

\begin{figure}[!tb]
\centering
\includegraphics[width=\columnwidth]{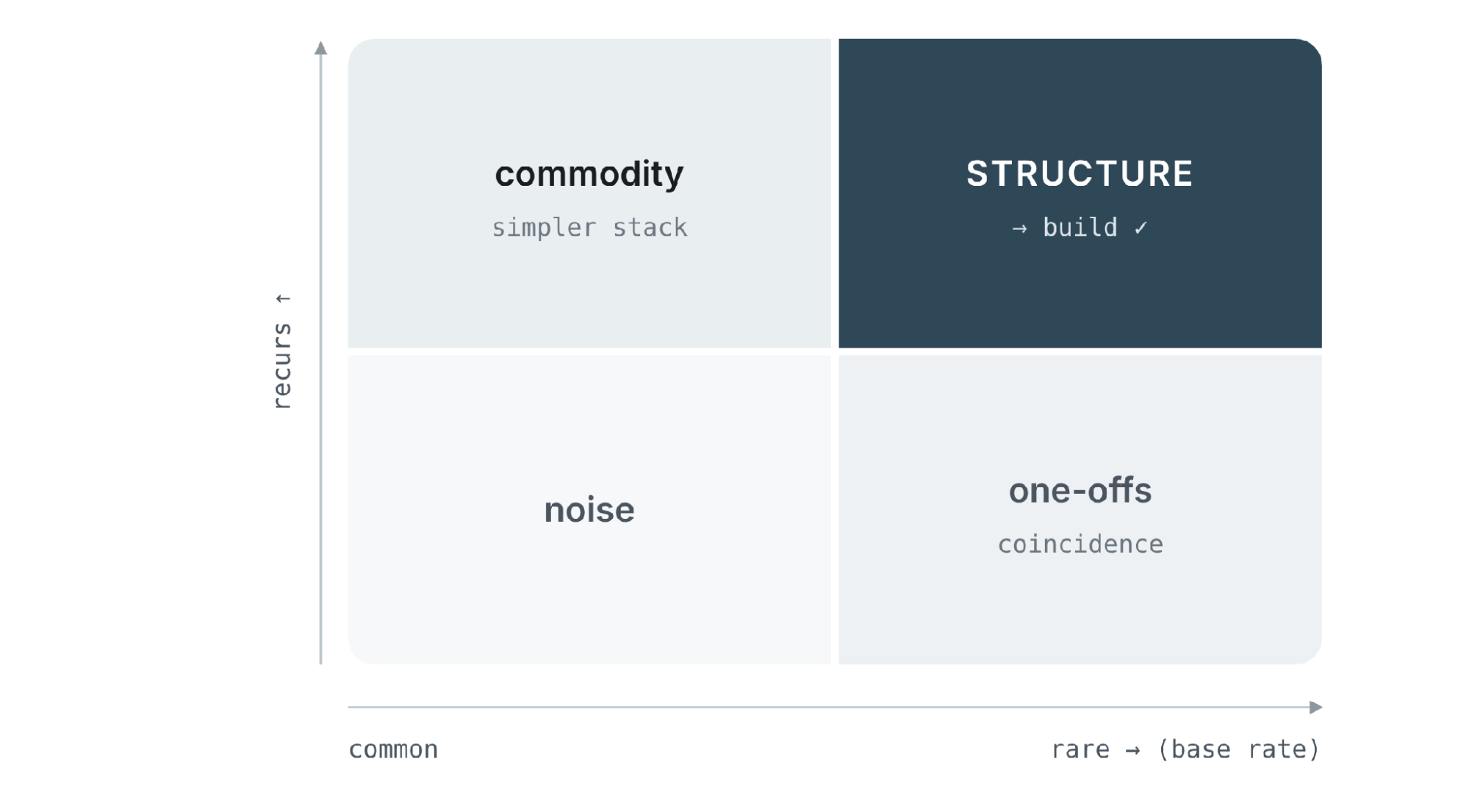}
\caption{The cheap probe: only links that both recur across related incidents and are rare in the background are worth the machinery.}
\label{fig:8}
\end{figure}

Figure~\ref{fig:8} is that grid. Only pieces in one corner, both recurring and rare, are the stable, distinctive structure a brain is meant to capture. If that corner is populated, there is something worth building. If the only links that recur are commodity, a simpler detection stack will do, and you have learned it in days.

\paragraph{Then build the smallest useful version.} Build the \textbf{schema} above first. Both the report-extractor and the telemetry pipeline must write that same object. Then build a single use, behavioral detection, delivered as a recommendation rather than a rule. It is the only use with a clean feedback loop, since an analyst confirms or dismisses each fire, and its clean links are what sharing and discovery would later stand on.

Sharing, discovery, and attribution come afterward. For the shape of a working minimum, see the reference implementation in appendix 3.

\paragraph{Rules that hold throughout.} Two commitments keep this safe to attempt. First, nothing on the path requires saying \emph{who}: attribution stays optional and last, and a brain claims a coherent pattern rather than a culprit. Second, an adversary can fake the traces used to name it, so naming is never settled and every brain stays a revisable hypothesis.

Three rules then hold permanently: keep each brain’s status visible (a full, weighted brain can look like a fact when it is a guess), let evidence flow only upward (a brain’s own conclusions never re-enter as new evidence), and keep naming decoupled from matching (one bad attribution, fed back in, poisons everything).

\section*{Appendix 1: Brain schema, version 0.1}
\addcontentsline{toc}{section}{Appendix 1: Brain schema, version 0.1}
\label{app:1}

This is the reproducible specification. A brain is one UTF-8 YAML document in one file, validated by a published JSON Schema. YAML is the authoring format because humans review it in pull requests, and JSON is the interchange format. The STIX export, specified below and not built, would produce it. The one schema validates both after parsing. Field names, allowed values, and sharing rules below are normative. Anything not listed is rejected by validation.

\begin{table*}[!t]
\centering
\caption{Brain object. Top-level document fields.}
\label{tab:1}
\small
\begin{tabularx}{\linewidth}{@{}>{\raggedright\arraybackslash}p{0.19\linewidth}>{\raggedright\arraybackslash}p{0.17\linewidth}>{\raggedright\arraybackslash}p{0.11\linewidth}>{\raggedright\arraybackslash}X@{}}
\toprule
\textbf{Field} & \textbf{Type} & \textbf{Required} & \textbf{Notes} \\ \midrule
\texttt{schema} & string & required & Fixed: \texttt{\textquotedbl{}brain/0.1\textquotedbl{}} \\
\texttt{id} & string & required & UUIDv4, stable for the brain’s life \\
\texttt{title} & string & required & Max 256 chars, sentence case \\
\texttt{status} & enum & required & experimental \textbar{} test \textbar{} stable \textbar{} retired \\
\texttt{scale} & enum & required & link \textbar{} technique-set \textbar{} attack-chain \textbar{} campaign \textbar{} threat-actor \\
\texttt{representation} & enum & required & graph \textbar{} list \textbar{} chain \textbar{} centroid \textbar{} embedding \\
\texttt{match} & enum & required & exact \textbar{} distance \textbar{} similarity \textbar{} sequence \textbar{} likelihood \\
\texttt{threshold} & number & optional & 0..1, default 0.7; see fire semantics \\
\texttt{tlp} & enum & required & clear \textbar{} green \textbar{} amber \textbar{} amber-strict \textbar{} red \\
\texttt{author} & string & optional &  \\
\texttt{created} & date & required & ISO 8601 (YYYY-MM-DD), quoted in YAML \\
\texttt{modified} & date & required & ISO 8601 (YYYY-MM-DD), quoted in YAML \\
\texttt{references} & list of strings & optional & URLs of source reports \\
\texttt{related} & list & optional & \{id: UUID, type: derived \textbar{} obsoletes \textbar{} merged \textbar{} split-from\} \\
\texttt{fidelity} & number & LOCAL & confirmed matches / total, set by the loop \\
\texttt{entities} & list of entity objects & required & min 2 \\
\texttt{links} & list of link objects & required & min 1 \\
\bottomrule
\end{tabularx}
\end{table*}

There is no attribution field, and the omission is deliberate. An attribution, that an actor is a particular organization, is a conclusion drawn from observables, not an observable itself, so no data can fill or confirm it, and a field the matcher reads must be one data can fill. A matcher that read it would use the conclusion to find its own supporting evidence, the circularity the whole system prevents. The name therefore lives in the title or references as prose, never as a matchable field.

\begin{table*}[!t]
\centering
\caption{Entity object. Declared once, referenced by links.}
\label{tab:2}
\small
\begin{tabularx}{\linewidth}{@{}>{\raggedright\arraybackslash}p{0.19\linewidth}>{\raggedright\arraybackslash}p{0.17\linewidth}>{\raggedright\arraybackslash}p{0.11\linewidth}>{\raggedright\arraybackslash}X@{}}
\toprule
\textbf{Field} & \textbf{Type} & \textbf{Required} & \textbf{Notes} \\ \midrule
\texttt{id} & string & required & Local ref, unique in file, e.g. \texttt{\textquotedbl{}e1\textquotedbl{}} \\
\texttt{type} & string & required & A STIX SCO type where one exists (domain-name, ipv4-addr, file, process, user-account, ...), else a namespaced custom type, e.g. \texttt{\textquotedbl{}x-\allowbreak{}vendor-\allowbreak{}thing\textquotedbl{}} \\
\texttt{value} & string & optional & A concrete value (\texttt{\textquotedbl{}evil.example\textquotedbl{}}) \\
\texttt{pattern} & map & optional & Field constraints when no concrete value, e.g. \texttt{\{name: \textquotedbl{}powershell.exe\textquotedbl{}, command\_line\textbar{}contains: \textquotedbl{}-enc\textquotedbl{}\}} \\
\bottomrule
\end{tabularx}
\end{table*}

{\footnotesize\noindent A key may end in an operator: \texttt{\textbar{}equals} (the default), \texttt{\textbar{}contains}, \texttt{\textbar{}starts}, \texttt{\textbar{}ends}, or \texttt{\textbar{}not-\allowbreak{}contains}. Keys starting \texttt{x-\allowbreak{}} are notes and are never matched. A pattern of only \texttt{x-\allowbreak{}} keys matches any entity of its type. Exactly one of value or pattern is required.\par\par}

The operators are close to Sigma field modifiers~\cite{r31}, which detection engineers already write, and like Sigma they ignore case in values. A STIX pattern~\cite{r33} could express most of these constraints, so a STIX export, specified below and not built, could carry a patterned entity as one. The schema keeps its own small set. It leaves out regular expressions, which STIX allows, because their cost is hard to bound in a shared file. STIX comparisons are case-sensitive.

Entity patterns cannot require separate observations to share a value, such as one account. In a brain that is the job of match: sequence, which binds an entity reused across links (fire semantics, below), and version 0.1 does not define its order fields. The flat rule does not bind, so an entity named in two links can match different values. STIX patterning has no such binding. It can require order (FOLLOWEDBY) and a time window (WITHIN). A STIX pattern is true or false, with no weights or threshold, so it cannot express the fire rule. Sigma correlations have it through group-by~\cite{r14}.

\begin{table*}[!t]
\centering
\caption{Link object. Two entities joined by one typed relation.}
\label{tab:3}
\small
\begin{tabularx}{\linewidth}{@{}>{\raggedright\arraybackslash}p{0.19\linewidth}>{\raggedright\arraybackslash}p{0.17\linewidth}>{\raggedright\arraybackslash}p{0.11\linewidth}>{\raggedright\arraybackslash}X@{}}
\toprule
\textbf{Field} & \textbf{Type} & \textbf{Required} & \textbf{Notes} \\ \midrule
\texttt{id} & string & required & Local ref, unique in file, e.g. \texttt{\textquotedbl{}L1\textquotedbl{}} \\
\texttt{type} & enum & required & state \textbar{} activity \\
\texttt{source} & string & required & An entity id \\
\texttt{target} & string & required & An entity id; may equal source for an action an entity takes on itself \\
\texttt{relation} & string & required & Verb from the recommended vocabulary (shared-registrant, resolves-to, spawned, connected-to, wrote, authenticated-to, ...) or a namespaced custom verb \\
\texttt{role} & enum & required & required \textbar{} supporting \\
\texttt{volatility} & enum & required & stable \textbar{} rotating \textbar{} ephemeral \\
\texttt{prior} & number & optional & 0..1, the sender’s suggested strength \\
\texttt{provenance} & map & required & \{kind: observed \textbar{} derived \textbar{} reported, source: free text or URL\} \\
\texttt{weight} & number & LOCAL & Computed against local base rates \\
\texttt{observability} & enum & LOCAL & visible \textbar{} invisible \\
\texttt{first\_seen} & date & LOCAL &  \\
\texttt{last\_seen} & date & LOCAL &  \\
\bottomrule
\end{tabularx}
\end{table*}

Fields marked LOCAL are stripped on publish and recomputed by every receiver: weight, observability, and fidelity are environment facts, and sightings expose the victim. A validator must reject a shared file that contains them.

\paragraph{Fire semantics.} A brain fires when every link with role required is matched and the weight-weighted share of matched supporting links meets its threshold. A brain needs at least one activity link to fire, which the reference implementation in appendix 3 does not yet check.

A required link whose observability is invisible suspends the brain rather than failing it. Invisible supporting links are left out of the share, and a brain with no visible supporting links fires on its required links alone. An entity that matches by itself adds to the score a pass uses to rank brains. Firing is defined over links.

This fire rule is flat: it cannot express “fire if any of several sub-cores is met,” so an actor whose campaigns share no common required core cannot be one brain that fires on any of them. Two paths are open and unbuilt: a disjunctive fire form over sub-cores, or the actor kept as a label grouping separate campaign brains, which the scale field already permits.

The rule also ignores order. Common pieces often appear together by chance, so a brain made of common pieces depends on order. \texttt{match: sequence} is for that case: it requires the links in order and for one actor, such as the same account or host. Version 0.1 does not yet define the fields that declare that order, or how centroid and embedding brains fire. Nor does it define where a brain’s links must co-occur. The reference implementation in appendix 3 uses one host within a fixed 24-hour window and matches state links across all hosts and times.

A detection compiled from a brain carries \texttt{compiled\_\allowbreak{}from: \textless{}brain id\textgreater{}}, and ingest drops any observation whose provenance chain contains a \texttt{compiled\_\allowbreak{}from} tag, which is how evidence flows only upward. Severity, how bad a fire is if it holds, is deliberately not a brain field. It turns on the asset hit and the stage reached, so it is worked out locally at match time and never shared.

\paragraph{Ingestion.} Each observation is committed to one frame on the way in:

\begin{itemize}

\item \textbf{IoCs and intel feeds.} Entities and state links: hashes, IPs, domains, registrants, certificates.

\item \textbf{Threat intel reports.} Extracted whole: entities plus state and activity links, arriving as weighted priors.

\item \textbf{Network telemetry.} Flows, DNS, proxy, and mail: activity events between entities.

\item \textbf{Endpoint and identity events.} Process, file, registry, and authentication: activity, normalized by an event schema such as OCSF~\cite{r32}.

\item \textbf{Control and SOC alerts.} Findings from DLP, email gateways, cloud controls, and detection fires. Second-order activity, another system’s judgment rather than a raw observation, carrying derived provenance.

\end{itemize}

\begin{figure*}[!tbp]
\begin{lstlisting}[caption={Repository layout.},label={lst:1}]
brains/
  link/              one file per brain: <slug>.yaml
  technique-set/
  attack-chain/
  campaign/
  threat-actor/
schema/
  brain-0.1.schema.json    JSON Schema used by CI
.github/workflows/
  validate.yml             on pull request: schema validation,
                           LOCAL-field rejection, id uniqueness,
                           related-id resolution
\end{lstlisting}
\end{figure*}

Contribution is by pull request. Merge requires validation to pass. A deployment may add human review. Status moves experimental to test to stable through review, or through the loop’s own confirmations where policy allows, and retired brains stay in history.

Status is the publisher’s assessment, where each receiver still earns its own trust in a brain, as local fidelity. For interchange beyond git, export maps entities to STIX cyber-observable objects, links to STIX relationship objects, and the brain to a bundle carrying its id and TLP marking~\cite{r33}\cite{r34}.

\begin{figure*}[!tbp]
\begin{lstlisting}[caption={Worked example. A minimal campaign-scale brain{,} valid against the schema{,} as published (no LOCAL fields).},label={lst:2}]
schema: brain/0.1
id: 6f1c2a9e-3d47-4b8a-9c11-2e5f8b7a4d01
title: Lure chain with registrant reuse
status: test
scale: campaign
representation: graph
match: likelihood
threshold: 0.7
tlp: green
created: "2026-07-01"
modified: "2026-07-08"
references:
  - https://example.org/report-2026-06
related:
  - {id: 4b0e8d21-77aa-4c55-8e02-91b3c6d0f2aa, type: derived}   # parent; must exist in this repository
entities:
  - {id: e1, type: domain-name, value: lure.example}
  - {id: e2, type: domain-name, value: stage.example}
  - {id: e3, type: process, pattern: {name: powershell.exe,
      command_line|contains: "-enc"}}
  - {id: e4, type: ipv4-addr, value: 203.0.113.7}
links:
  - id: L1
    type: state
    source: e1
    target: e2
    relation: shared-registrant
    role: required
    volatility: rotating
    prior: 0.8
    provenance: {kind: reported, source: report-2026-06}
  - id: L2
    type: activity
    source: e3
    target: e4
    relation: connected-to
    role: required
    volatility: ephemeral
    prior: 0.6
    provenance: {kind: observed, source: edr}
  - id: L3
    type: activity
    source: e2
    target: e3
    relation: delivered
    role: supporting
    volatility: rotating
    prior: 0.5
    provenance: {kind: reported, source: report-2026-06}
\end{lstlisting}
\end{figure*}

\begin{figure*}[!tbp]
\centering
\includegraphics[width=0.55\textwidth]{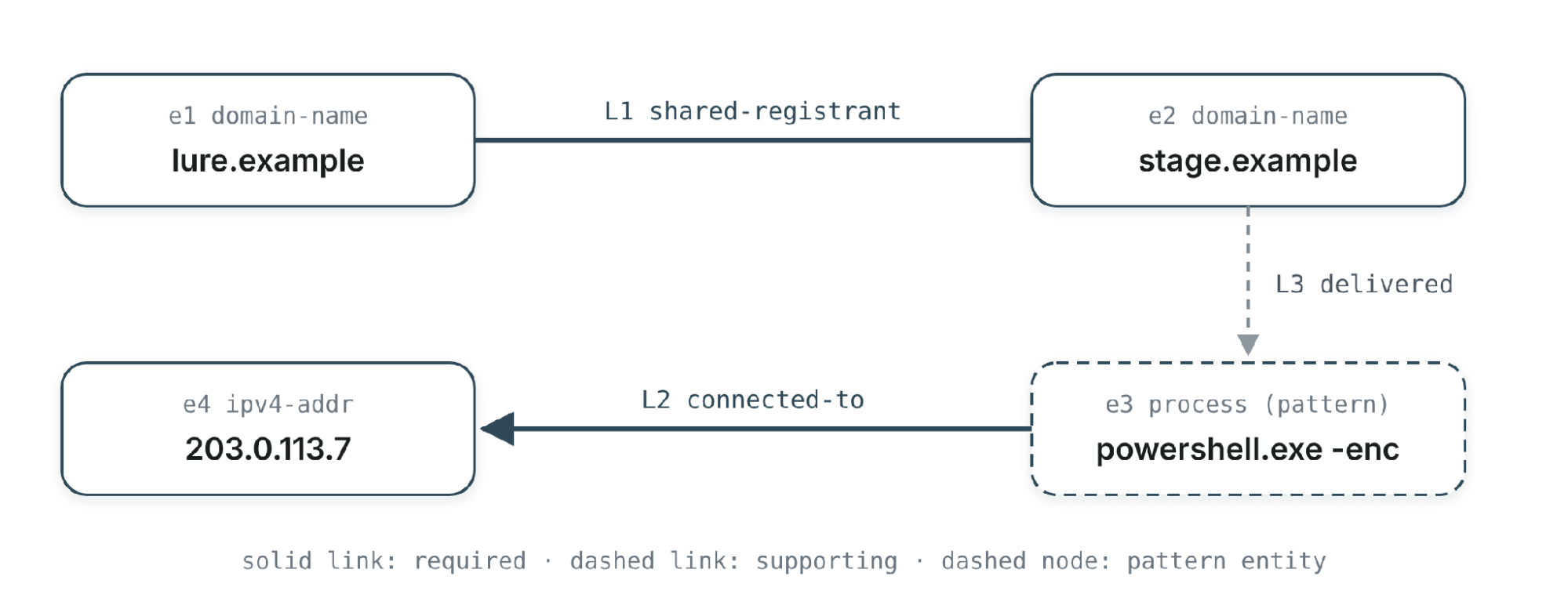}
\caption{The worked example, drawn. Listing~\ref{lst:2} is one serialization of this graph: entities are the nodes, links the edges, with L1 and L2 required and L3 supporting.}
\label{fig:9}
\end{figure*}

\section*{Appendix 2: Building a brain from a report}
\addcontentsline{toc}{section}{Appendix 2: Building a brain from a report}
\label{app:2}

This appendix runs the Extract operation and the wager test on real public reports. It works through five along one axis, how specific the actor’s structure is: from a single rare structural link that lets a brain alert on its own, through behaviors that only rank, to an actor with no distinctive piece, where the brain is only an arrangement or does not form.

How specific a brain is decides what it can do. Each case shows the field choices it forces, with a brain rendered in proportion to its specificity: near-complete where one link carries it, a sample scaffold where only the arrangement matters. Every technical detail is drawn from the cited source.

\paragraph{APT1, the clean case.} Mandiant’s 2013 report gives a rare, distinctive behavior and a large pile of indicators, and the two age very differently. The behavior: an APT1 beachhead backdoor (WEBC2) retrieves a web page from a C2 server and interprets the data between special HTML tags as commands. Older variants read the data from HTML comments, and later variants moved it into other tags~\cite{r6}.

Mandiant had collected more than 400 samples of WEBC2 variants, released over more than six years, the earliest compiled in January 2004. Alongside the behavior, the report published over 3,000 indicators: domains, addresses, and hashes. Three months later Mandiant reported that APT1 had stopped using the vast majority of that infrastructure and shifted to new tools and infrastructure~\cite{r5}.

That split decides the build: the indicators become entities, and any link through them is marked \texttt{ephemeral}: real, but retooled within months, so they carry little weight. The command-retrieval behavior becomes one \texttt{activity} link marked \texttt{required} and \texttt{stable}: it is the habit the group is named for, and it survived six years of variants.

The report’s further claim, that APT1 is PLA Unit 61398, is the attribution the schema keeps out of matching: the brain represents APT1 as an actor-scale pattern and can carry that name. Only the identification with an organization stays in the title and references.

\begin{figure*}[!tbp]
\begin{lstlisting}[caption={The APT1 brain.},label={lst:3}]
schema: brain/0.1
id: 7d3f1a02-9c4e-4b21-a7f0-2e9b5c81d004
title: APT1 command retrieval in web-page tags   # name in the title, not a field
status: stable
scale: threat-actor
representation: graph
match: likelihood
threshold: 0.7
tlp: clear
created: "2026-07-01"
modified: "2026-07-08"
references:
  - https://services.google.com/fh/files/misc/mandiant-apt1-report.pdf
entities:
  - {id: e1, type: process, pattern: {name: webc2-backdoor}}   # family label; needs a normalizer
  - {id: e2, type: url, pattern: {x-role: c2-webpage}}
  - {id: e3, type: file, pattern: {x-note: spearphish-zip-attachment}}
  - {id: e4, type: domain-name, value: aaa.aaa.example}   # one of 3000+; retooled in months
links:
  - id: L1
    type: activity
    source: e1
    target: e2
    relation: retrieves-commands-in-html
    role: required
    volatility: stable          # 400+ samples over six years; the namesake habit
    prior: 0.9
    provenance: {kind: reported, source: mandiant-apt1-2013}
  - id: L2
    type: activity
    source: e3
    target: e1
    relation: delivered
    role: supporting
    volatility: rotating
    prior: 0.4
    provenance: {kind: reported, source: mandiant-apt1-2013}
\end{lstlisting}
\end{figure*}

The report also documented HTRAN hop-forwarding, observed in 767 instances, but that traffic is visible from a research vantage, not from a typical victim network. Added as a link, it would be \texttt{invisible} for most receivers: as a required link it would suspend the brain, and as a supporting link it would drop out of the supporting share. The brain is left with one strong, stable link and a cloud of dead indicators. One durable habit can outlast a large indicator set, and a brain like this ranks rather than alerts, since its link is seen only through a malware classifier.

\paragraph{SUNBURST, where the strong link is structural.} The SolarWinds backdoor first checks that its own host process is \texttt{solarwinds.businesslayerhost}, then beacons out. The distinctive fact is that the signed Orion monitoring process itself is making the outbound call~\cite{r7}. That is not a tool choice the operator can swap without giving up the operation, so it is \texttt{required} and \texttt{stable}, and it is observable on the host. Its command-and-control domain, avsvmcloud.com, was seized and turned into a killswitch within days of disclosure~\cite{r35}, so a link through it would be \texttt{ephemeral} and carry little weight.

\begin{figure*}[!tbp]
\begin{lstlisting}[caption={The SUNBURST brain{,} entities and links.},label={lst:4}]
entities:
  - {id: e1, type: process, pattern: {name: solarwinds.businesslayerhost.exe}}
  - {id: e2, type: domain-name, pattern: {value|not-contains: solarwinds,
                                          x-note: non-solarwinds-destination}}
links:
  - id: L1
    type: activity
    source: e1
    target: e2
    relation: connected-to
    role: required
    volatility: stable          # the monitoring process itself beacons; cannot be swapped away
    prior: 0.9
    provenance: {kind: reported, source: fireeye-sunburst-2020}
\end{lstlisting}
\end{figure*}

Unlike APT1, the required link here is rare, structural, and observable, so the brain can alert on its own once it earns high fidelity, provided the destination test separates the beacon from Orion’s ordinary outbound traffic. On our synthetic background a substring test alone did not separate them (appendix 3).

The links worth the most are the ones tied to what the operation is, not to its current tooling and infrastructure, and a burned indicator drops to low weight while the structural link keeps the brain alive.

\paragraph{Scattered Spider, where the signature is off-telemetry.} The group’s entry is a phone call to a help desk to reset multi-factor authentication~\cite{r36}. That step is the deciding one, and it is not in any telemetry a brain ingests, so receivers mark its link \texttt{invisible} and the brain suspends for want of visibility.

What can be built is a scaffold of observable steps that are each routine but distinctive in arrangement: registering a new MFA device, then adding a federated identity provider to the single sign-on tenant with automatic account linking, then changing conditional-access policies to widen access~\cite{r36}\cite{r37}.

CISA records the identity-provider step as a historical tactic, still possibly in use but no longer confirmed as current~\cite{r36}, so its link is marked \texttt{rotating}. CISA notes that the actors often change TTPs and rotate machine names~\cite{r36}, so links through indicators are ephemeral.

\begin{figure*}[!tbp]
\begin{lstlisting}[caption={The Scattered Spider scaffold{,} entities and links.},label={lst:5}]
entities:
  - {id: e1, type: user-account, pattern: {x-note: help-desk reset target}}
  - {id: e2, type: x-identity-provider, pattern: {x-note: newly federated, x-auto-link: "on"}}
  - {id: e3, type: x-help-desk, pattern: {x-note: help-desk channel, x-channel: phone}}
links:
  - {id: L0, type: activity, source: e3, target: e1, relation: reset-mfa-via-helpdesk,
     role: required, volatility: stable, prior: 0.8,   # off-telemetry in most environments
     provenance: {kind: reported, source: cisa-aa23-320a}}
  - {id: L1, type: activity, source: e1, target: e1, relation: registered-mfa-device,
     role: required, volatility: stable, prior: 0.6,
     provenance: {kind: reported, source: cisa-aa23-320a}}
  - {id: L2, type: activity, source: e2, target: e1, relation: federated-with-auto-link,
     role: required, volatility: rotating, prior: 0.7,   # historical tactic, no longer confirmed current
     provenance: {kind: reported, source: cisa-aa23-320a}}
  - {id: L3, type: activity, source: e1, target: e1, relation: widened-access-policy,
     role: supporting, volatility: stable, prior: 0.5,
     provenance: {kind: reported, source: microsoft-octo-tempest-2023}}
\end{lstlisting}
\end{figure*}
\begin{figure*}[!tbp]
\begin{lstlisting}[caption={The Volt Typhoon brain{,} entities and links.},label={lst:6}]
entities:
  - {id: e1, type: process, pattern: {name: netsh.exe, command_line|contains: portproxy}}
  - {id: e2, type: process, pattern: {name: ntdsutil.exe}}
links:
  - {id: L1, type: activity, source: e1, target: e1, relation: created-port-proxy,
     role: required, volatility: stable, prior: 0.8,
     provenance: {kind: reported, source: cisa-aa24-038a}}
  - {id: L2, type: activity, source: e2, target: e2, relation: copied-directory-database,
     role: required, volatility: stable, prior: 0.8,
     provenance: {kind: reported, source: cisa-aa24-038a}}
\end{lstlisting}
\end{figure*}
\begin{figure*}[!tbp]
\begin{lstlisting}[caption={The Akira technique-set brain{,} entities and links.},label={lst:7}]
entities:
  - {id: e1, type: process, pattern: {x-note: ESXi encryptor}}
  - {id: e2, type: file, pattern: {name|ends: .akira}}   # a later ESXi variant uses .akiranew
links:
  - {id: L1, type: activity, source: e1, target: e2, relation: wrote,
     role: required, volatility: stable, prior: 0.7,     # scale: technique-set
     provenance: {kind: reported, source: cisa-aa24-109a}}
\end{lstlisting}
\end{figure*}

The arrangement is the signature here, since no single step is rare. The build clears the wager only as an attack-chain scaffold, and only where the identity events are logged. Where the deciding link is the phone call, the response is to improve help-desk verification and telemetry.

\paragraph{Volt Typhoon, an actor built to be featureless.} The group lives off the land, using built-in tools and valid credentials to avoid leaving distinctive artifacts, and relays its traffic through a botnet of home routers~\cite{r38}. Most of what it does is commodity by design.

Two links survive the wager: a \texttt{netsh} PortProxy registry change that turns a compromised host into a relay, which Microsoft’s published hunting query matches on the specific command combining \texttt{portproxy}, \texttt{netsh}, \texttt{wmic}, \texttt{process call create}, and v4tov4~\cite{r39}, and ntdsutil writing a copy of the directory database to a readable location~\cite{r38}.

Both are observable, and both are close to the operation rather than to a swappable tool, so they are \texttt{required} and \texttt{stable}. The router relays are infrastructure: the botnet was disrupted by a US operation~\cite{r38}, so links through those entities are \texttt{ephemeral}. Even an actor built to be featureless leaves a few native actions it cannot avoid, and a small behavioral brain forms around them, while the actor’s infrastructure burns and its behavior links persist.

\paragraph{The ransomware floor, where the wager fails.} RansomHub is ransomware-as-a-service, run by many affiliates~\cite{r40}, and both its advisory and Akira’s read as commodity throughout: initial access through an edge-device or VPN vulnerability, credential theft with Mimikatz, volume-shadow-copy deletion, and double extortion~\cite{r40}\cite{r41}.

The addresses in RansomHub’s advisory overlap with earlier QakBot infrastructure, and CISA advises vetting them before blocking~\cite{r40}. That is, they do not discriminate. An actor brain built from either report would match thousands of unrelated intrusions.

The right answer is to build nothing at that scale, and the wager test reaches it in days. Where a distinctive piece exists, it is the tool, not the actor: Akira’s ESXi encryptor writing its own file extension (.akira, or .akiranew in a later version) is a narrow technique-set brain worth keeping~\cite{r41}. Otherwise a single rule on the exploited edge vulnerability does the same work as a brain, for far less. A brain that never forms is the common and correct outcome.

\paragraph{What the five cases show together.} A report yields a brain worth building when it contains a piece that is rare, durable, and observable at once, and that combination is scarcer than the volume of most reports suggests. SUNBURST has it in a single structural link, and APT1 nearly so, in one habit seen only through a malware classifier. Volt Typhoon has it in two native actions, while Scattered Spider has it only in an arrangement, and only where the identity layer is logged. The ransomware floor, in the observed data, does not have it at the actor scale at all.

Each field earns its place across these cases. Weight tracks how rare a piece is in ordinary traffic, volatility tracks how fast it changes, observability decides whether it can be seen here, and fidelity, earned later from confirmed matches, decides whether the brain alerts or only ranks. Read as brains, the most-cited reports are mostly small: one durable link and a large perishable remainder. So specificity sets the fidelity a brain can earn, and fidelity sets use, and the wager is whether the actor left anything specific enough to cross the alert line.

\section*{Appendix 3: The reference implementation}
\addcontentsline{toc}{section}{Appendix 3: The reference implementation}
\label{app:3}

We built the schema of appendix 1 as a small engine, to test its rules. It ships six brains, five from the appendix 2 cases and the worked example, each one YAML file that passes validation. Its Scattered Spider brain keeps an earlier version of link L3, with relation \texttt{deleted-\allowbreak{}access-\allowbreak{}policy}, because the synthetic tests and their recorded results use that name.

The engine applies the fire rule, suspension, a deterministic rarity screen, the loop, and Discover. Most tests run on synthetic data we generated, so they test mechanisms and give no detection rate.

Where the engine uses a language model, rules that run outside the model check its output. To judge a match, the model gets a match block: the pieces that matched deterministically, each with a rarity score. Evidence that names none of these pieces is discarded. Confidence is capped at 0.4 unless one of these pieces has a rarity of 0.5 or more. Extraction drops any technique id that is not in a small curated subset of ATT\&CK. Without a model endpoint, the engine replays recorded model output. The tests of these rules use no model.

\paragraph{Limits of the reference implementation.} Compile and Share are not built, so the \texttt{compiled\_from} rule is not exercised. There are no centroid or embedding brains. Drift flagging, running a brain in reverse, Coverage-map’s three-way sort, and alert-or-rank routing by fidelity are not built. The engine reads a supplied list of invisible links. The screen is not yet connected to the model, so the tests supply the match block.

The engine stores representation, match, prior, and volatility but does not use them to match. Order is checked by a separate sequence matcher, over required links only, used in one test. The loop’s verdicts come from the synthetic data generator’s labels.

\begin{itemize}

\item \textbf{Order.} On 6,000 synthetic host-days, a flat rule over two common pieces fired on all 6,000. On a separate synthetic set of 6,000 host-days, three common identity steps, required in order and for one user, fired on 5, a lower bound because the matcher is greedy. The same steps out of order, or spread across three users, gave only a partial match. A Sigma temporal\_ordered correlation grouped by host and user, with a 24-hour timespan, expresses nearly the same rule~\cite{r14}, and so can an EQL sequence~\cite{r15}. Both slide their time window, where the engine uses a fixed host-day.

\item \textbf{One link and several.} The one-link SUNBURST brain fired on 594 of 6,000 host-days. Its destination test is a substring check, and the background includes Orion connecting to non-SolarWinds hosts at a rate we set. The multi-link brains, and the APT1 and Akira brains, each with one required link, fired on none. Their pieces are absent from, or rare in, the background we set. A C2 domain that contains \texttt{\textquotedbl{}solarwinds\textquotedbl{}} evades the SUNBURST brain.

\item \textbf{Separation.} Under a null model of independent pieces, the Volt Typhoon brain, with two required links, separated from chance, and the Scattered Spider brain, with three required links of which one is common, narrowly did not.

\par What decided separation was how rare the links were. Real pieces co-occur, so this test is optimistic. In a separate run that checked fire rates only, a correlated background left them nearly unchanged (SUNBURST 582 of 6,000, against 561 with independent pieces, and none for the multi-link brains).

\item \textbf{Rotation and loss.} Over 100 runs, 99 with rotated indicators, the behavioral brain fired every time. A twin built on the indicator fired on the first run only. Removing the event for a supporting link moved a brain from fired to partial. A required link marked invisible suspended the brain.

\item \textbf{The loop.} A brain confirmed 30 of 30 times reached fidelity 1.0, and a brain never confirmed in 566 fires was retired. Split and merge kept lineage, and weights did not change as brains were added, split, or merged. Discover found the planted pattern in 30 of 30 runs, as it must when a pattern recurs often enough, and also proposed about 25 benign patterns per run, which is the work disproof would need to prune.

\item \textbf{Weight and the loop.} The engine computes weight as the negative log, base 2, of a link’s rate in a synthetic background, kept apart from library rarity. The screen computes library rarity as one over the number of brains that name a piece. In synthetic tests, a contradiction lowers a brain’s standing and an absence does not. Disproof is not built.

\item \textbf{One emulated case.} On public emulated data for an lsass memory dump through comsvcs.dll~\cite{r42}, a brain written for this technique caught the one attack and fired on none of seven other techniques. A Sigma rule did the same. This run used the flat rule. When the command line was changed, the brain still fired and a command-line Sigma rule missed. A Sigma rule on the same call trace would still fire, and a dump through dbghelp instead of dbgcore defeats the brain. One case is not a rate.

\end{itemize}

\paragraph{Durability in public reporting.} We built 25 brains from the first report in each of eight vendor series on six actors. For five series, we checked later reports for each required link, and a recurrence counted only where the actor did it again and a later report observed it independently.

Of 47 verdicts, 23 could be judged: 9 showed the link persisting and 4 showed it replaced by a successor, all from ongoing operations (Scattered Spider, Volt Typhoon, Akira). The other 10 did not count toward durability: 7 reflected analysts learning more, 2 a vendor reframing its own finding, and 1 showed a RansomHub link fading. For SUNBURST and RansomHub, no recurrence counted.

The three rotating links that later reports addressed all changed. Two Scattered Spider links authored as stable, both tied to Azure, were later replaced, by 2025 with VMware techniques. Reports are a biased sample, so a link that does not recur may still be durable.

Turning these prose-built brains into matchers left 7 of 71 entities as prose that could not be compiled to a match rule, which is the work Compile has to do. The study also points to three loop moves the text does not yet specify: linking a faded brain to its successor (which \texttt{related: obsoletes} can record), changing a brain’s scale, and tightening a brain in place. The last two rest on one case each.

\end{document}